# Planetary thermal evolution models with tectonic transitions

Craig O'Neill

Macquarie Planetary Research Centre

Macquarie University, Sydney, Australia.

Thermal history calculations provide important insights into the interior evolution of planets, but incorporate simplified dynamics from the systems they represent. Planetary interiors typical incorporate complex rheologies, viscous layering, lateral heterogeneities, and time delays in processes, which have not been traditionally represented by parameterised approaches. Here we develop numerical models for planetary evolution, incorporating the physical complexity of Earth's interior, and use them to generate statistically-based Nu-Ra scalings. These encapsulate the main effects of tectonic transitions, geometry, and depth-dependent rheology, and time-sensitivity. We find an exponent $\beta$ of ~0.26 best describes the Nu-Ra relationship for evolving mobile lid systems, and $\beta$ ~0.12 for stagnant-lid systems. Systems with time dependent subduction have $\beta$ varying between ~0.26 during the Hadean, when external factors such as impacts facilitate tectonics, to ~0.12 during the Archaean, when the system is dominated by long periods of quiescence, and systems driven by external forcings (eg. due to impacts in the first 100Myr of Earth's history) may exhibit much higher exponents. We also find a time-lag between Ra (which primarily depends on mantle temperature) and Nu (normalised surface heat flow) of around 200-300Myr, suggesting a significant delay between mantle thermal configuration, and its surface manifestation. These results provide an approach for the rapid characterisation of tectonic, volcanic, and atmospheric evolution of terrestrial exoplanets.

# Introduction

Thermal evolution models offer a unique approach to understanding the evolution of temperatures within planetary interiors, and constraining their volcanism and surface geology. Typically thermal evolution models adopt a parameterised approach to mantle convection, utilising a well-known relationship between convective vigour (the Rayleigh number - Ra), and normalised heat flow, generally cast as the ratio of convective to conductive heat, a term called the Nusselt number (Nu; Christensen, 1984). Boundary layer theory (Turcotte and Oxburgh, 1967), laboratory experiments (Booker, 1976) and numerical experiments (eg. Christensen, 1984) have consistently found that these two terms follow a relationship of the form:

$$Nu \approx \alpha Ra^\beta$$

(Equation 1)

Boundary layer analysis suggests that β is around 1/3, a result confirmed by numerical simulations of simple (isoviscous) systems, and α is generally determined experimentally. However, for more rheologically complex systems, β may vary. For example Christensen (1984) found that β was around ~0.1, for stagnant-lid systems with high-viscosity lids. Gurnis (1989) emphasised that for systems with mobile lids (ie. viscous lids, with weak zones to permit plate motions), a more traditional scaling of β ~ 1/3 is appropriate. Moresi and Solomatov (1998) found that β may be between 0.224-0.293 for mobile-lid systems with yielding, depending on how internal viscosity is defined.

Variations in the aspect ratio of the flow can also impact heat transport (Hansen and Ebel, 1984; Olson, 1987), decreasing the efficacy of heat loss, and affecting primarily β in the Ra-Nu scaling. Channelised mantle flow, due to vertical mantle viscosity variations, affects α for similar reasons (eg. Lenardic et al., 2006). These factors are likely to be relevant to the Earth. The Ra-Nu systematics of an Earth-like system, with viscosities determined by mineral physics, and flow in a spherical-shell, have not been adequately constrained.

Furthermore, the possibility of tectonics transitions within Earth's history (eg. O'Neill et al., 2016) may significantly perturb Earth's thermal history (Silver and Behn, 2008; Korenaga, 2013). As an example, parameterised models integrating backwards, and assuming a value of β ~ 1/3 typically predict mantle temperatures of > 2000K within the last 1-3Gyr. This would imply a primarily molten, very-high temperature mantle throughout the Archaean which is not born out by the volcanic record (Silver and Behn, 2008; Korenaga, 2006). Dubbed *'The Archaean thermal catastrophe'* this behaviour is primarily due to the efficiency of heat loss under a hard β ~ 1/3 mobile scaling. If heat was lost in the past as efficiently as plate tectonics today, then energy conservation requires that the interior was hotter to provide this heat flux. Factors postulated to avoid this pathological behaviour include altering the Urey ratio (the ratio of internal heating to surface heat flux), enhanced slab strength and bending resistance (Conrad and Hager, 2001; Korenaga, 2006), and the possibility of tectonic transitions (O'Neill et al., 2013) or periods of plate-tectonic cessation (Silver and Behn, 2008) decreasing the heat loss efficiency in the past.

These and related problems require a revision of the Ra-Nu scaling for Earth-like systems under different tectonic regimes. Whilst modern numerical convection simulations are capable of addressing this problem, they are restricted by computational overheads from exploring large parameter spaces, and their adoption somewhat limited by steep learning curves for individual researchers, due to complex nature of even mature community codes. A parameterised approach, however, can cover a wide parameter space over billions of years of evolution efficiently, and their conceptual ease lends them to adoption across diverse disciplines (including geological, astronomical and exoplanet communities). Our approach here is to utilise advanced numerical simulations of tectonic systems on an evolving Earth-like planet with a complex rheology, to refine a Nu-Ra relationship using a brute-force statistical approach, to aid, as an example, exoplanet thermal evolution models.

## Method

*Mantle convection*

To simulate the dynamics and evolution of Earth-like planets, we use the community mantle convection code *Aspect* (aspect.dealii.org) to solve the coupled set of equations for mantle convection (conservation of mass, momentum (ie. Stokes equation) and energy), respectively:

$$\nabla \cdot \vec{u} = -(\frac{1}{\rho}\frac{\partial \rho}{\partial p})\rho g \vec{u}$$

(Equation 2)

$$-\nabla \cdot \left[2\eta \left(\dot{\varepsilon}(\vec{u}) - \frac{1}{3}(\nabla \cdot \vec{u})I\right)\right] + \nabla p = \rho \vec{g}$$

(Equation 3)

$$\rho C_p \left(\frac{\partial T}{\partial t} + \vec{u} \cdot \nabla T\right) - \nabla \cdot k\nabla T = \rho H + \alpha T(\vec{u} \cdot \nabla p) + 2\eta \left(\dot{\varepsilon}(\vec{u}) - \frac{1}{3}(\nabla \cdot \vec{u})I\right) : \left(\dot{\varepsilon}(\vec{u}) - \frac{1}{3}(\nabla \cdot \vec{u})I\right)$$

(Equation 4)

Here the velocity is $\vec{u}$, density $\rho$, pressure $p$, gravitational acceleration $g$, viscosity $\eta$, strain rate tensor $\dot{\varepsilon}(\vec{u}) = \frac{1}{2}(\nabla\vec{u} + \nabla\vec{u}^T)$, heat capacity $Cp$, thermal conductivity $k$, internal heating rate $H$, and thermal expansivity $\alpha$. Compressibility is implemented as vertically compressible only. Note in Equation 4 the heat source H decays through time. Over 2 million grid cells are used, in an axisymmetric 2D plane (see Figure 1), and the computational mesh is refined near the surface. Our top and bottom boundary conditions are free slip, the top set at a constant temperature (293K), and the bottom at a temperature determined by an evolving core model (see O'Neill et al., 2017, for details).

The viscosity we use incorporates dislocation, diffusion and Peierl's creep mechanisms, and is determined by:

$$\eta_{diff/disl/P} = A^{-\frac{1}{n}}\exp\left(\frac{E + pV}{nRT}\right)\dot{\varepsilon}_{II}^{\frac{1-n}{n}}$$

(Equation 5)

R is the gas constant, A is a prefactor, n the stress exponent, E the activation energy, V the activation volume, and $\dot{\varepsilon}_{II}$ the second invariant of the strain rate tensor.

We also include yielding via an effective viscosity mechanism of the form:

$$\eta_y = \frac{\tau_0 + fp}{2\dot{\varepsilon}_{II}}$$

(Equation 6)

Where the surface yield strength is $\tau_0$, and the friction coefficient $f$. The composite viscosity is then calculated from

$$\eta = 1/(\frac{1}{\eta_{diff}} + \frac{1}{\eta_{disl}} + \frac{1}{\eta_P} + \frac{1}{\eta_y})$$

(Equation 7)

We have included the full input file for our simulations in the supplementary sections, and the full listing of the parameters used in our models is shown in Table 1.

| Earth Radius | R (km) | 6371 |
|---|---|---|
| Core Radius | $R_c$ (km) | 3481 |
| Gravity acceleration | g (m/s$^2$) | 9.81 |
| Initial CMB temperature | $T_{CMB}$ (K) | 4400 (varies) |
| Core density | $\rho_c$ (kg/m$^3$) | 12.5×10$^3$ |
| Core heat capacity | $Cp_c$ (J/kg/K) | 840 |
| Core thermal expansivity | $\alpha_c$ (K$^{-1}$) | 1.35×10$^{-5}$ |
| Core thermal conductivity | $K_c$ (W/m/K) | 40 |
| Pre-factor, upper mantle diffusion creep | A (Pa$^{-n}$s$^{-1}$) | 3.0×10$^{-11}$ |
| Pre-factor, lower mantle diffusion creep | A (Pa$^{-n}$s$^{-1}$) | 2.2×10$^{-15}$ |
| Pre-factor, dislocation creep | A (Pa$^{-n}$s$^{-1}$) | 5.0×10$^{-19}$ |
| Pre-factor, Peierls creep | A (Pa$^{-n}$s$^{-1}$) | 1×10$^{-150}$ |
| Activation energy, upper mantle diffusion creep | E (kJ/mol) | 300 |
| Activation energy, lower mantle diffusion creep | E (kJ/mol) | 200 |
| Activation energy, dislocation creep | E (kJ/mol) | 540 |
| Activation energy, Peierls creep | E (kJ/mol) | 540 |
| Activation volume, upper mantle diffusion creep | V (cm$^3$/mol) | 4 |
| Activation volume, lower mantle diffusion creep | V (cm$^3$/mol) | 1.5 |
| Activation volume, dislocation creep | V (cm$^3$/mol) | 12 |
| Activation volume, Peierls creep | V (cm$^3$/mol) | 10 |
| Stress exponent, upper mantle diffusion creep | n | 1 |
| Stress exponent, lower mantle diffusion creep | n | 1 |

| Stress exponent, dislocation creep | n | 3.5 |
|---|---|---|
| Stress exponent, Peierls creep | n | 20 |
| Surface yield strength | $\tau_0$ (MPa) | 10-50 (20 default) |
| Friction coefficient | f | 0.1-0.5 (0.2 default) |

*Table 1. Description of parameters for the numerical simulations and material properties. Note depth-dependent parameters (mantle density, thermal capacity, thermal conductivity, etc.) are calculated from Perplex lookup tables (see (http://www.perplex.ethz.ch/) using the thermodynamic database of Stixrude & Lithgow-Bertelloni (2011) and assuming a pyrolitic composition. Mantle is assumed to melt beyond the solidus (defined by Stixrude & Lithgow-Bertelloni (2011)), leaving the mantle capped at solidus temperatures.*

*Parameterised model*

The starting point for a parameterised model of mantle convection is a global energy conservation equation of the form:

$$C_{Earth}\frac{dT}{dt} = H(t) - Q(t)$$

(Equation 8)

Here $C_{Earth}$ is the heat capacity of the Earth (~$7\times10^{27}$ J/K; Korenaga, 2003), and following Christensen (1984) and Korenaga (2003) T is the mantle potential temperature, H is the average heat production of the bulk silicate Earth (BSE), and Q is the surface heat loss. The heat production through time is given by

$$H = \sum_i C_0^i H^i exp\left(\frac{tln2}{\tau_{1/2}^i}\right)$$

(Equation 9)

Here i refers to the radioactive species ($^{238}$U, $^{235}$U, $^{40}$K or $^{232}$Th), C the initial isotope concentration, t is time, and $\tau_{1/2}$ its half-life. These parameters are listed in Table 2, and H in Equation 8 would be scaled to W for the mantle's mass ($4\times10^{24}$ kg).

| Isotope | H (W/kg) | $\tau_{1/2}$ (yr) | C (kg/kg) |
|---|---|---|---|
| $^{238}$U | 9.46 x 10$^{-5}$ | 4.47 x 10$^9$ | 30.8 x 10$^{-9}$ |
| $^{235}$U | 5.69 x 10$^{-4}$ | 7.04 x 10$^8$ | 0.22 x 10$^{-9}$ |
| $^{232}$Th | 2.64 x 10$^{-5}$ | 1.40 x 10$^{10}$ | 124 x 10$^{-9}$ |
| $^{40}$K | 2.92 x 10$^{-5}$ | 1.25 x 10$^9$ | 36.9 x 10$^{-9}$ |

*Table 1: Heat producing isotopes and their properties, based on Turcotte and Schubert (2002).*

To integrate Equation 8 forward in time, we apply an initial mantle temperature, and first determine the average mantle viscosity using a simplified Arrhenius form:

$$\eta = A_0 \exp\left(\frac{E}{RT}\right)$$

(Equation 10)

This form is used for the parameterised models as they do not track stress/strain, unlike the numerical models, and thus a Newtonian is required. Here $A_0$ is determined empirically, E is 300 kJ/mol, and T is updated each timestep. Based on this average viscosity, one can determine a system Rayleigh number:

$$Ra = \frac{\alpha \rho g \Delta T d^3}{\kappa \eta}$$

(Equation 11)

Here $\alpha$, $\rho$ and g are mantle averages (ie. including pressure effects), and are 1.5x10$^{-5}$ K$^{-1}$, 4500 kg/m$^3$ and 9.81 m/s$^2$ respectively. $\Delta T$ is the temperature difference between the surface (T = 0°C) and the average interior potential temperature $T_p$ (ie. T minus the adiabatic contribution), and d is the mantle thickness (2891 km). The thermal diffusivity $\kappa$ is 1x10$^{-6}$ m$^2$/s, and viscosity $\eta$ is from Equation 10. Equation 1 then converts Ra into Nusselt number, which can be converted into global heat flux Q by multiplying by the conductive heat, given by kAΔT /d, where k is thermal conductivity (4.5 W/m.K), and A the surface area of the globe (given by $4\pi r^2$). Q and H in Equation 8 are now determined, and one can calculate the change in temperature dT, and update the internal temperature accordingly.

Lastly, previously scalings have emphasised the effect of the viscosity contrast (θ) of the mantle, which moderates Eq 1 by an additional factor of $\theta^{-\gamma}$ (Moresi and Solomatov, 1995), where γ is a fitted exponent. We note in our simulations that the effective viscosity contrast is capped, by a constant surface temperature, and also internal temperatures restricted by the solidus. As a result, the viscosity contrast across the system is generally constant.

# Results

We initially consider the evolution of an Earth-like convecting system, with complex rheologies, viscosity layering, and evolving thermal conditions. We have performed over 20 simulations from different starting conditions, yield strength parameters, impact flux, and tracked the evolution of Ra and Nu dynamically in each (see Supplementary Figure S1). An example time series from one of our mantle convection simulations is shown in Figure 1. The model evolves forward in time, with heat production decreasing, and core temperatures dropping according to a core evolution model. The models also have a flux of impactors perturbing the thermal state of the mantle, as per the details given in O'Neill et al. (2017). These are most significant in the first 100Myr, transitioning to acting as a trigger mechanism for ongoing tectonic resurgences after this. After the initial pulse of impact-related activity, the reference model evolves into a stagnant-lid state, interspersed with periods of occasional activity, for the first 600Myr.

The time evolution of Ra and Nu for this system are shown in Figure 2 for this period, and Figure 3 for the interval 700-1200Myr (ie. ~3.85 - 3.35Ga in Earth's history). The Ra is calculated from the average mantle viscosity away from the boundary layers, and core temperature, and depth averaged values for the other terms, which are assumed constant, as given in Equation 11. The Nusselt number is calculated from the geometrically scaled heat flux (ie. scaled by $4\pi r^2/(2\pi r)$), divided by the (spherical) conductive component, incorporating the radiogenic contributions for the whole mantle volume ($H_r$), given by:

$$Q_{conductive} = \frac{4\pi k \Delta T r_1 r_2}{r_2 - r_1} + H_r$$

From about 700Myr on, the models evolve into an episodic state, with intermittent bursts of tectonic activity interspersed with periods of relative stasis. The latter state is not a true stagnant lid - the lid in these cases exhibits significant internal deformation, but is not strictly mobile, nor does it exhibit wholesale recycling, except during episodic subduction episodes.

In contrast to this, Figure 1 also exhibits some timeslices from a true mobile lid model, which has the same configuration and system evolution as the previous example, but with the yield stress dropped to ensure lid mobility (ie. $\tau_0$=10MPa and f=0.1 in Equation 6, see Table 1).

For the mobile case, the Ra-Nu relationship implies a β of ~0.26 (Figure 4), in line with earlier estimates of β from Moresi and Solomatov (1998), for simple Cartesian steady-state mobile systems. Note the scaling coefficient α includes geometrical effects, and the fit in Figure 4 is not directly applicable to 3D systems, as it will vary for 2 vs 3D (eg. due to channelised flow, Lenardic et al., 2006). The earliest period of these models (t < 100Myr) exhibit both start-up effects, and the predominance of forced mobile overturns (due to large impacts). The exponent in this effectively "driven" regime is quite high (0.41, similar in mobile and "quasi-stagnant" models), indicating a level of heat loss well above that expected from boundary layer theory, due to external drivers. It also highlights the significant affect large impacts may have planetary cooling histories.

In Figure 2c we show 4 different Ra-Nu fits, for different time periods of the one model shown in Figure 1, including the average for the whole period (green), which is strongly influenced by start-up effects. The average for the period t > 100Myr is shown in black, and exhibits a β of ~0.26, which is very similar to the nominally mobile-lid case. The model here exhibits periods of lid-overturn, often triggered by impact events, and as a result approaches mobile-lid in heat-loss efficiency. Also shown are scalings for when the mobility (the ratio of $V_{Surface}/V_{RMS}$ (Tackley 2000)) is both greater than 1.1 (blue) and < 0.9 (red). The former approaches TBL theory in its efficiency, though of course surface velocities and internal temperatures are far above the mean for this system. The low mobility periods are likewise affected by enhanced cooling between quiescent episodes, and so these scalings are not independent.

The period for which the episodic reference model was continued into the Archaean (Figure 3) shows a low value of β of ~0.12. This is very low, and approximates values found by Christensen (1984) for stagnant lid models. The model, throughout most of the period, exhibits stagnant lid behaviour, with occasional bouts of subduction, and this is reflected in the large spread of heat flux (Nu) values, and mantle velocities (eg. through Ra) shown in Figure 3c. This also represents an evolution of the system away from the effects of "forced" tectonics (ie. impact effects in the Hadean), to a more volcanogenic Archaean regime, and low heat loss rates may represent the system equilibrating from its earlier over-cooling.

Lastly, Figures 2d-4d exhibit correlation functions between the natural logs of Ra and Nu. The plots indicate significant correlation between Ra and Nu, but imply a significant lag time (245-305 Myr for the episodic model, 195 Myr for the mobile model), implying a sizeable delay between surface heat loss and mantle temperature equilibration. This lag time is of the order of the internal heating timescale of the mantle (eg. Piccolo et al., 2019). These timescales have analogues in the 'reactance times' of Lenardic et al. (2019), the time it took for a perturbation to decay in their parameterised models. They found models with low β had longer reactance times, resulting in unstable runaway perturbations if β < 0. Our lag times, however, arise from the finite thermal adjustments times of numerical models, and are quantitatively different, although they behave qualitatively similar.

## Discussion and Conclusions

The results demonstrate that planetary mantles are capable of exhibiting a range of β over their history, in response to the evolution of their tectonic regimes, and also external factors (such as bombardment-forcings). Variable β may mitigate some of the issues faced by thermal evolution models assuming a single value. One particular case in the point is the Archaean thermal catastrophe, which, as alluded to by Silver and Behn (2008), is largely an artefact of assuming current plate tectonic scalings are appropriate over all of Earth's history. If we relax this constraint and permit the tectonic transitions over time (O'Neill et al., 2016), this catastrophe is largely alleviated.

An example is shown in Figure 5, of three end-member parameterised models evolving from an identical initial temperature. This is a fairly simplistic simulation that ignores variations in relative heat flow at the CMB for instance (ie. it has a constant Urey number), or divergent layered mantle evolution, but is used to demonstrate the difference the tectonic scalings derived here may have on simple systems, and is similar in approach to some of the key past examples (eg. Christensen, 1984; Korenaga, 2003). The initial temperature is set at 2073 K (near the peridotite solidus). The lines represent the evolution of a stagnant-lid model ($\beta$=0.12, red dashed line), a strict mobile-lid model ($\beta$=0.26, blue dashed line), and a hybrid regime which alternates over the course of its history, from $\beta$=0.26 in its 0.5Gyr (ie. the Hadean), to $\beta$=0.12 from 0.5 to 2.250 Gyr (roughly the Archaean) to a more traditional mobile-lid scaling ($\beta$=0.26) from 2.25-4.5Gyr (roughly the Proterozoic on). Shown also are the mantle temperature estimates from volcanic rocks of Herzberg et al. (2010, black dots, and KDE).

The mobile lid model overcools early in its history, and largely underestimates mantle temperatures for much of the past. The pure stagnant-lid model does not cool effectively enough, and exhibits hot (up to solidus) mantle temperatures for its entire evolution. The hybrid model has imposed transitions at major geological boundaries in Earth's history. For instance the transition from significant impact activity, to primarily tectonic and volcanic activity, has been argued to take place around the Hadean-Archaean boundary (Marchi et al., 2014, O'Neill et al., 2017), and the transition to plate tectonics has been suggested to occur around the late Archaean (O'Neill et al., 2018, and references therein). With these transitions imposed, a fairly sensible thermal history can be constructed, that satisfies implied initial conditions (O'Neill et al., 2017), paleo-temperature estimates (Herzberg et al., 2010) and current mantle temperatures, without requiring unusual material behaviour (Conrad and Hager, 2000).

One restriction on this current scaling is that we haven't explicitly considered volcanic heat transport. Moore and Webb (2013) showed that for volcanic heat pipe regimes with eruptive efficiencies of ~100% (ie. 100% of melt formed in the mantle is erupted extrusively), volcanic heat transport can dominate conduction through the lithosphere. However, more recent work (Rozel et al., 2017) considering melt intrusion within the crust and lithosphere has shown that for moderate eruption efficiencies of 20%, which is appropriate for many intraplate settings (Crisp et al., 1984), then most volcanic heat goes into heating the crust. Under those circumstances heat flows will approach traditional convective scalings. In addition it is not clear whether current eruption efficiencies are appropriate in the past. Many high-temperature melts observed in the Archaean (eg. komatiites) are dense due to the Mg-content (van Thienen et al. 2004) and may be intrusively emplaced, and so the ratio of erupted:intruded volcanic products may evolve over time. An extension of our current results into volcanic-dominated systems will require a resolution of these ambiguities.

Another restriction to our parameterised model is that we are not explicitly calculating core heat flux here (in contrast to our numerical simulations). This implicitly assumes the core is equilibrating to mantle temperatures - a behaviour seen in the numerical simulations - but which may lead to issues when this coupling does not hold. An example might be the early cooling of a superheated core, or the transition of predominant cooling mode of the core (eg. thermal to thermochemical). Parameterisations exist to explicitly model these interactions (eg. Stacy and Loper, 1984).

Extension of these simulation to 'super-Earth' sized exoplanets requires some consideration of the mantle layering and heat flow properties of such planets (eg. Tackley et al., 2013). In the case of extreme layering, it has been shown for the Earth that dramatic differences in thermal history are possible (McNamara and van Keken, 2000; Honda, 1995). However, this extreme layering probably has not applied to Earth over its history (McNamara and van Keken, 2000; O'Neill et al., 2013, 2018). The dynamic layering of superEarths will depend on their exact mineralogy, but is not expected to impede whole-mantle convection (Tackley et al., 2013), and thus similar surface scalings to those derived above should hold, albeit with modified viscosity descriptions.

Ultimately, the goal of such exoplanet interior modelling is to develop quantitative predictions that may be able to feed into exoplanet observations. This in large part means being able to predict diagnostic atmospheric compositions that are a function of tectonic regime. The classic example of this is plate tectonics on Earth, where the $CO_2$-silicate weathering cycle (a function of continual plate-boundary driven topography) removes atmospheric $CO_2$ (Sleep and Zahnle, 2001). Plate subduction also is able to recycle atmospheric $CO_2$ into the interior via buried carbonates. As a result, an Earth-like terrestrial exoplanet with an ocean-continent topographic dichotomy, may be expected to be atmospheric $CO_2$ poor. This simple relationship however would break down for a water-world or ice-world, where these weathering or burial processes are not operative, and other atmospheric constituents, such as the $N_2$ cycle, may be more informative (Zerkle and Mikhail, 2017).

Another example is the presence of oxygen. Whilst $O_2$ is a product of life on Earth, its accumulation in the atmosphere was delayed by up to 1Gyr - which has been argued to be due to extreme volcanic degassing prior to ~2.4Ga (Kump et al., 2001). So a simple prediction of a young volcanically and tectonically active planet may be, paradoxically, low atmospheric $O_2$. The counterpoint - a young planet with a rich $O_2$ atmosphere, may be in a stagnant-lid regime.

Other more spectroscopically challenging gas species, such as noble gases (and their isotopes), have the potential to be more diagnostic. For example, previous work (O'Neill et al., 2014) on Venus has suggested that the isotopic ratios of noble gases in the atmosphere of Venus may be used to constrain its history of degassing, and thus tectonic history. In the case of Venus, it has lost around 25% of the $^{40}$Ar that it should have produced over its history to the atmosphere (Earth has around 50%). This observation suggests neither planet has efficiently degassed, and it has been argued to support the case of Earth having plate tectonics over the last half of its history, and Venus having an early stagnant lid regime, transitioning to episodic tectonics, for most of its history (O'Neill et al., 2014). Such scenarios, though, are hampered by interior models of Venus (or Earth's) viscosity, and mantle heterogeneity.

The development of active tectonics in mantle simulations involves resolving small developing rheological and thermal heterogeneities in 2 or 3D, and this is a challenge to 1D parameterised models. An additional challenge to understanding the tectonics of exoplanets is that these tectonic transitions are sensitive to the thermal evolution of a planet (Crowley and O'Connell, 2012; Weller and Lenardic, 2012, O'Neill et al., 2016), and thus it is not sufficient to solely consider an exoplanet's physical state (size, density etc), its history is critical too. The scalings presented here allow for an examination of end-member thermal evolutionary histories for exoplanets, for different imposed tectonic regimes. The ultimate characterisation of exoplanet atmospheres will soon provide critical constraints on exoplanet volcanism, and tectonic evolution. Our results provide an approach to constraining the thermal evolutionary history of exoplanets based on future atmospheric observations.

## Acknowledgements

The author would like to thank Lena Noack and an anonymous reviewer for helpful comments.

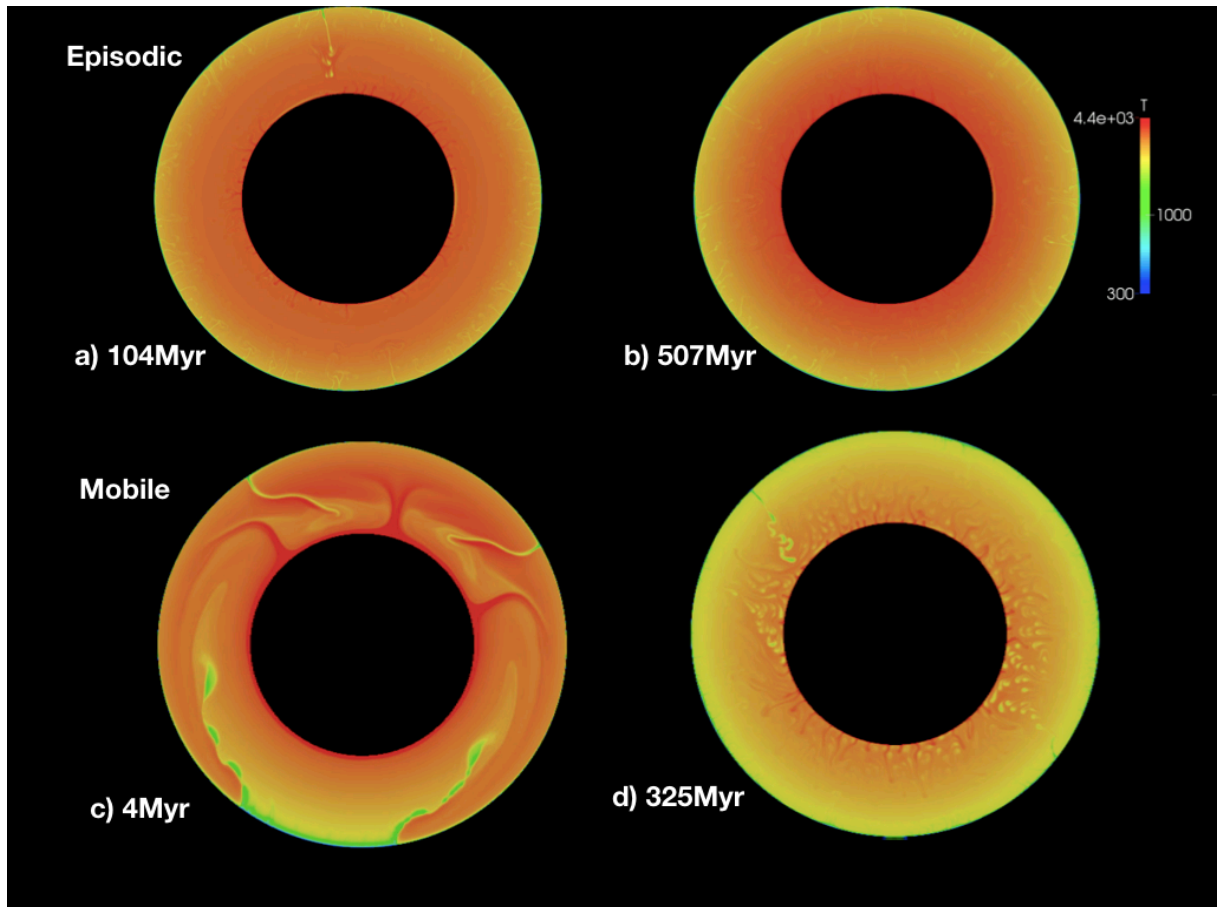

Figure 1. Snapshots of temperature field for two Earth-like convecting systems, in either an ostensibly episodic regime (a, b) or a mobile lid regime (c, d) for different times.

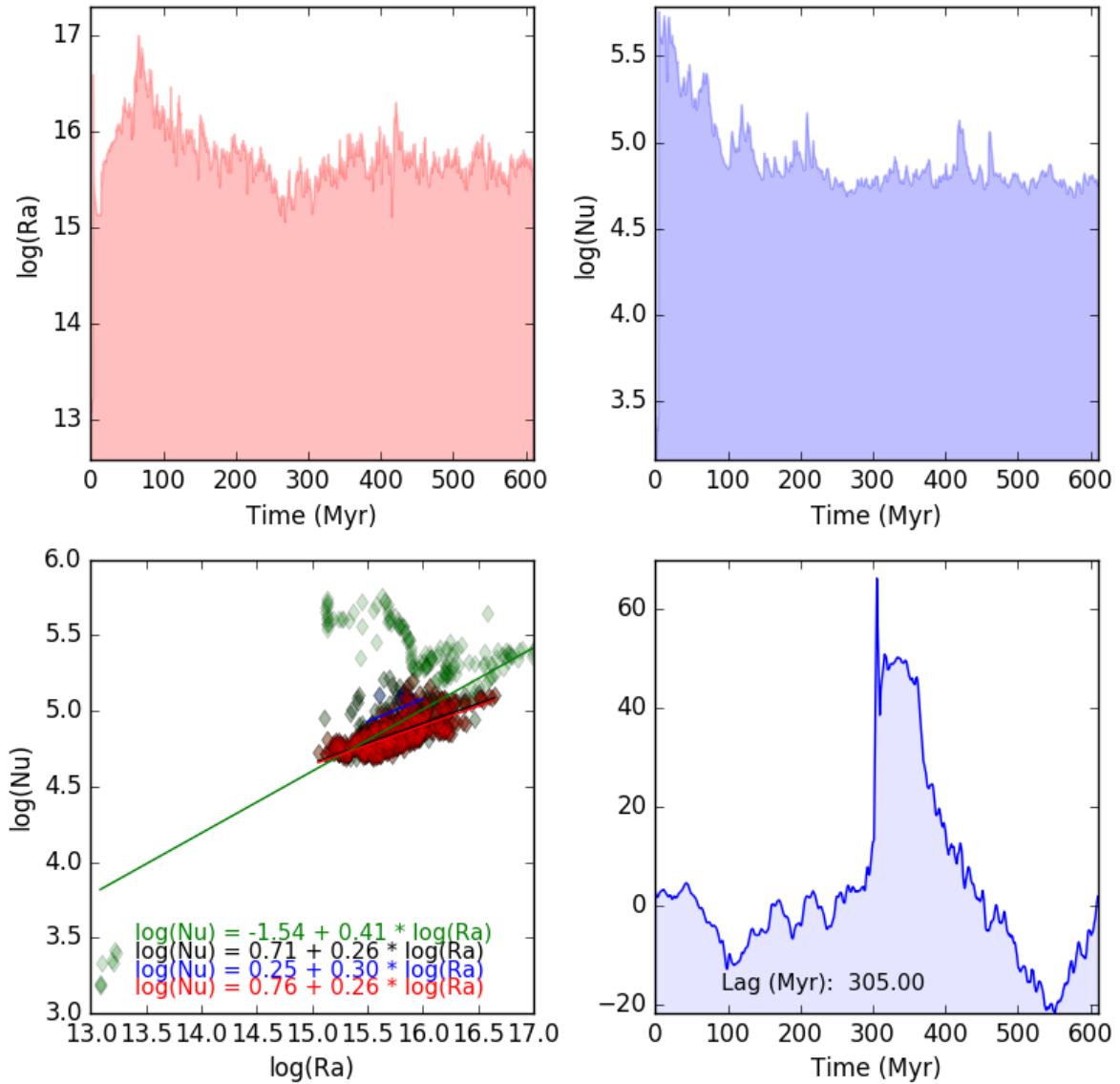

Figure 2. Evolution of the Ra (a), and Nu (b) for the 'episodic' convecting system shown in Figure 1 a and b. c) Plot of discretised natural log(Ra) vs log(Nu). The gradient of the fit defines β in equation 1. Green fit is for the entire run, black is restricted to t>100Myr, blue is for t>100Myr and mobility > 1.1, and red for t>100Myr and mobility < 0.9 (the latter two are not independent). d) Correlation function between log(Ra) and log(Nu), indicating a significant lag time (305 Myr) between the two time-series.

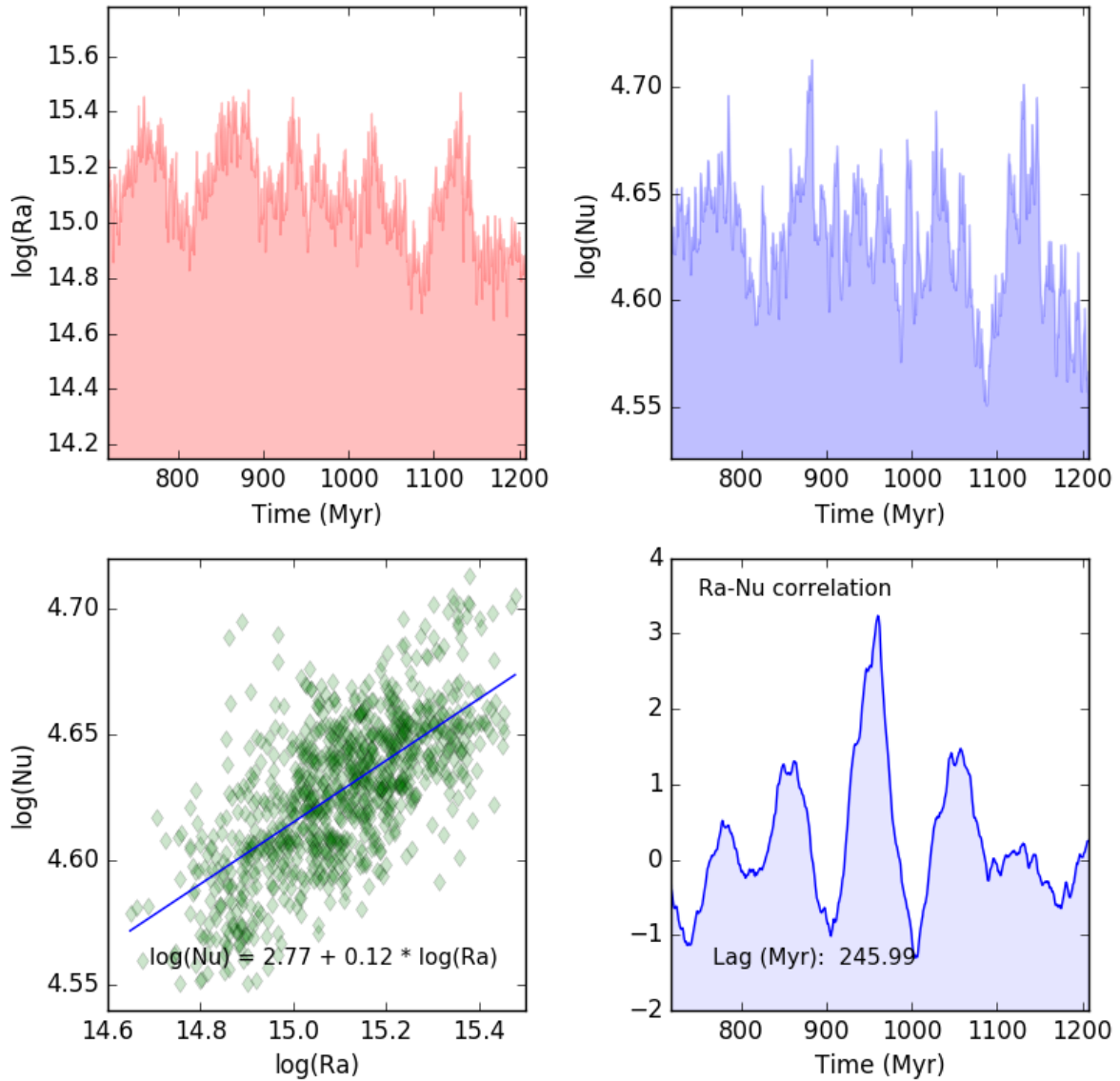

Figure 3. Evolution of the Ra (a), and Nu (b) for the 'episodic' convecting system shown in Figure 1 a and b, for times between 700-1200Myr. c) Plot of discretised natural log(Ra) vs log(Nu), for this time period. The gradient β is 0.12, consistent with long periods of stagnant lid activity. d) Correlation function between log(Ra) and log(Nu), indicating a significant lag time (245.99Myr) between the two time-series.

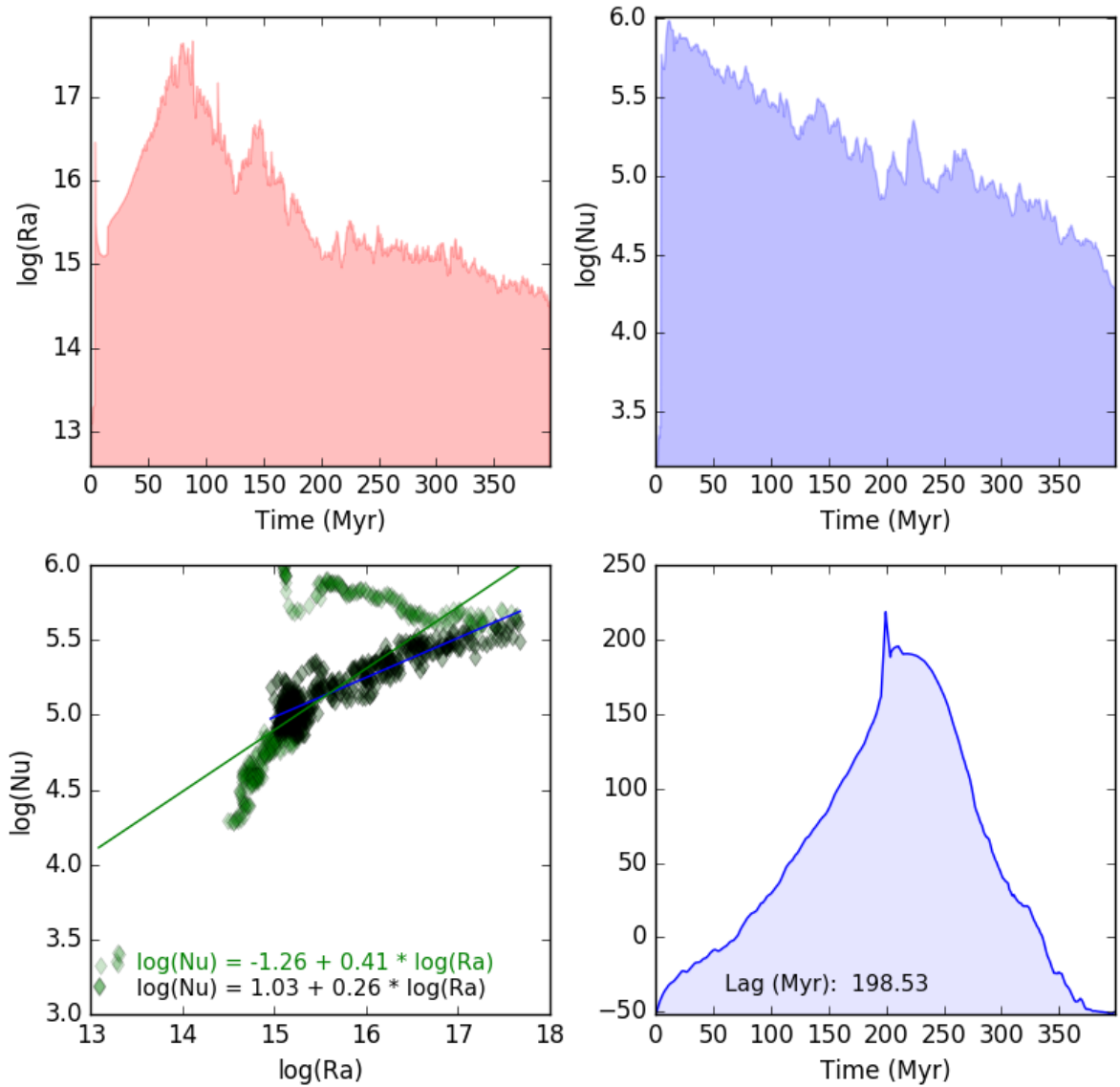

Figure 4. Evolution of the Ra (a), and Nu (b) for the 'mobile' convecting system shown in Figure 1 c and d. c) Plot of discretised natural log(Ra) vs log(Nu). Green fit is for the entire run, black fit (blue line) is restricted to t>100Myr. d) Correlation function between log(Ra) and log(Nu), indicating a significant lag time (198.53Myr) between the two time-series.

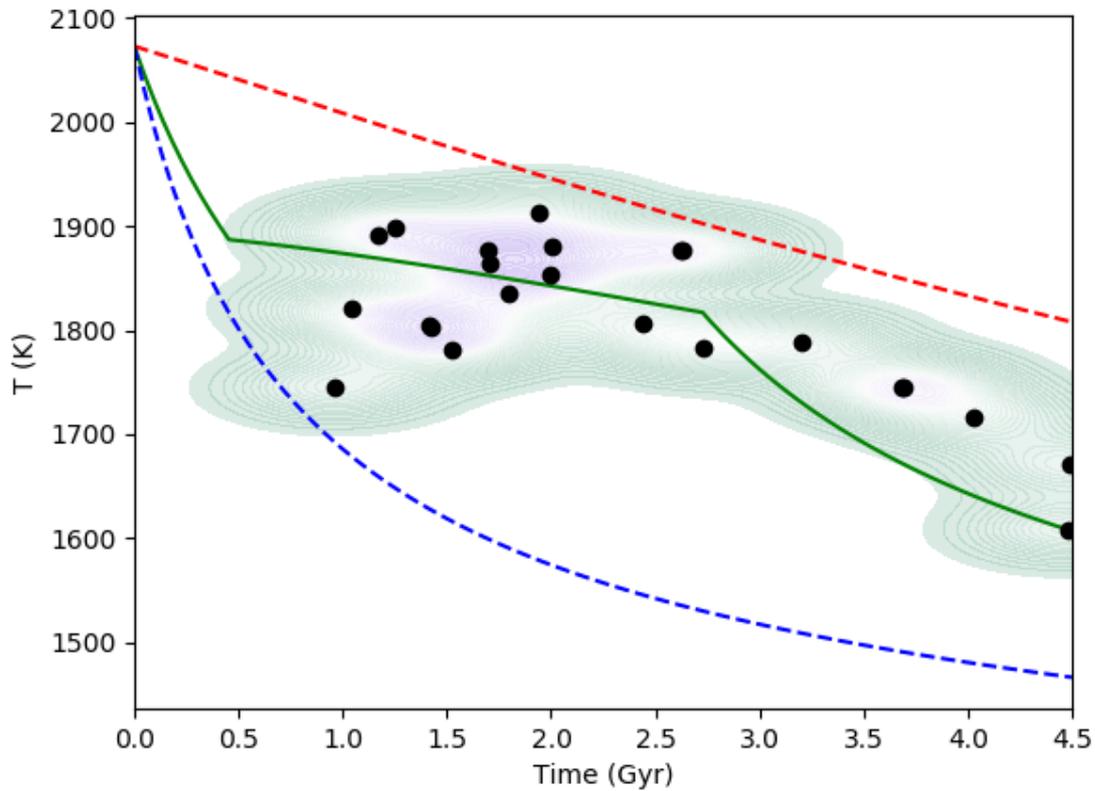

Figure 5. Evolution of mantle temperatures for 3 models from the same initial temperature (2073K), following different tectonic evolution trajectories. The blue dashed lines shows the evolution of a purely mobile-lid system (β=0.26), the red dashed line a purely stagnant-lid system (β=0.12), and the green line a system with tectonic transitions, evolving from an 'early episodic' regime with β=0.26, to 'quiescent-episodic' regime, with β=0.12, into a modern mobile-lid regime (β=0.26) at 2.0Ga. Black dots indicated temperature measurements from Herzberg et al. (2009), and coloured region indicates the kernel density distribution of this temperature data.